\documentclass[runningheads]{llncs}
\usepackage[T1]{fontenc}
\usepackage{amsmath,amssymb,amsfonts}
\usepackage{graphicx}
\usepackage{textcomp}
\usepackage{graphicx}
\usepackage{algorithm,algorithmic}
\usepackage{hyperref}
\usepackage{textcomp}
\usepackage{tabularx}
\usepackage{booktabs}

\usepackage{graphicx}
\begin{document}
\title{Structurally Different Neural Network Blocks for the Segmentation of Atrial and Aortic Perivascular Adipose Tissue in Multi-centre CT Angiography Scans}
\titlerunning{PVAT Segmentation}
%
\author{Ikboljon Sobirov$^{1,2,*}$, 
Cheng Xie$^{2*}$,
Muhammad Siddique$^{2,4}$,
Parijat Patel$^{2,4}$,
Kenneth Chan$^{2}$,
Thomas Halborg$^{2}$,
Christos P. Kotanidis$^{2}$,
Zarqaish Fatima$^{3}$,
Henry West$^{2}$,
Sheena Thomas$^{2}$, 
Maria Lyasheva$^{2}$, 
Donna Alexander$^{5}$, 
David Adlam$^{5}$, 
Praveen Rao$^{5}$, 
Das Indrajeet$^{5}$, 
Aparna Deshpande$^{5}$, 
Amrita Bajaj$^{5}$, 
Jonathan C L Rodrigues$^{6}$, 
Benjamin J Hudson$^{6}$, 
Vivek Srivastava$^{7}$, 
George Krasopoulos$^{7}$, 
Rana Sayeed$^{7}$,
Qiang Zhang$^{7}$,
Pete Tomlins$^{4}$,
Cheerag Shirodaria$^{4}$,
Keith M. Channon$^{2}$,
Stefan Neubauer$^{2}$,
Charalambos Antoniades$^{2}$,
and Mohammad Yaqub$^{1}$}
\authorrunning{I. Sobirov et al.}
%
\institute{$^{1}$ Department of Computer Vision, Mohamed bin Zayed University of Artificial Intelligence, Abu Dhabi, UAE\\
$^{2}$ Acute Multidisciplinary Imaging \& Interventional Centre, Division of Cardiovascular Medicine, Radcliffe Department of Medicine, University of Oxford, Oxford, UK \\
$^{3}$ Oxford University Hospitals NHS Foundation Trust, Oxford, UK \\
$^{4}$ Caristo Diagnostics LTD, Oxford, UK \\
$^{5}$ Department of Cardiovascular Sciences and NIHR Leicester Biomedical Research Centre, University of Leicester, Leicester, UK\\
$^{6}$ Department of Radiology, Royal United Hospitals Bath NHS Foundation Trust, Bath, UK\\
$^{7}$ Department of Cardiothoracic Surgery, Oxford, UK}

\maketitle              
\def\thefootnote{*}\footnotetext{These authors contributed equally to this work}
\begin{abstract}
Since the emergence of convolutional neural networks (CNNs) and, later, vision transformers (ViTs), deep learning architectures have predominantly relied on identical block types with varying hyperparameters. We propose a novel block alternation strategy to leverage the complementary strengths of different architectural designs, assembling structurally distinct components similar to Lego blocks. We introduce \emph{LegoNet}, a deep learning framework that alternates CNN-based and SwinViT-based blocks to enhance feature learning for medical image segmentation. We investigate three variations of \emph{LegoNet} and apply this concept to a previously unexplored clinical problem: the segmentation of the internal mammary artery (IMA), aorta, and perivascular adipose tissue (PVAT) from computed tomography angiography (CTA) scans. These PVAT regions have been shown to possess prognostic value in assessing cardiovascular risk and primary clinical outcomes. We evaluate \emph{LegoNet} on large datasets, achieving superior performance to other leading architectures.
Furthermore, we assess the model’s generalizability on external testing cohorts, where an expert clinician corrects the model’s segmentations, achieving DSC > 0.90 across various external, international, and public cohorts. To further validate the model’s clinical reliability, we perform intra- and inter-observer variability analysis, demonstrating strong agreement with human annotations. The proposed methodology has significant implications for diagnostic cardiovascular management and early prognosis, offering a robust, automated solution for vascular and perivascular segmentation and risk assessment in clinical practice, paving the way for personalised medicine.
\end{abstract}
\section{Introduction}
\label{sec:introduction}

From the early convolutional neural network (CNN)-based U-Net~\cite{cciccek20163d} to the most recent vision transformer (ViT) models~\cite{hatamizadeh2022unetr,hatamizadeh2022swin}, deep learning (DL) segmentation architectures follow the typical style of an encoder and decoder network, where the encoder is typically consists of a series of identical blocks with varying hyperparameters. This design is not limited to segmentation but extends to other tasks, such as classification and detection. While such architectures have demonstrated strong performance across various applications, little attention has been given to exploring alternative encoder designs that move beyond identical block structures. This raises a fundamental question: \textit{Does a deep learning encoder learn better representations when built with identical or non-identical blocks?}


We study the impact of harmonizing internally nonidentical blocks for segmenting the internal mammary artery (IMA), aorta, and perivascular adipose tissue (PVAT) from multi-centre computed tomography angiography (CTA) scans. While previous works have explored hybrid architectures that integrate ViT and CNN encoders~\cite{chen2021transunet,zhang2021transfuse}, either side-by-side or sequentially, to the best of our knowledge, no study has examined the block-level integration of different deep learning architectures. We propose an approach where structurally distinct yet compatible blocks are alternated within a deep learning model. This perspective introduces new possibilities in model design and block selection, which we evaluate using three types of blocks: CNN-based and SwinViT-based, resulting in three architectural variations. Conceptually, this approach resembles assembling a model using compatible Lego pieces, inspiring the name \emph{LegoNet}. We hypothesize that incorporating structurally diverse blocks can lead to richer feature representations, particularly in complex tasks like medical image segmentation. To validate this, we assess \emph{LegoNet} in the challenging task of vessel-level segmentation in 3D scans.

\begin{figure*}[t!]
\centering
\begin{tabular}{ccc}
{\includegraphics[width=0.3\textwidth]{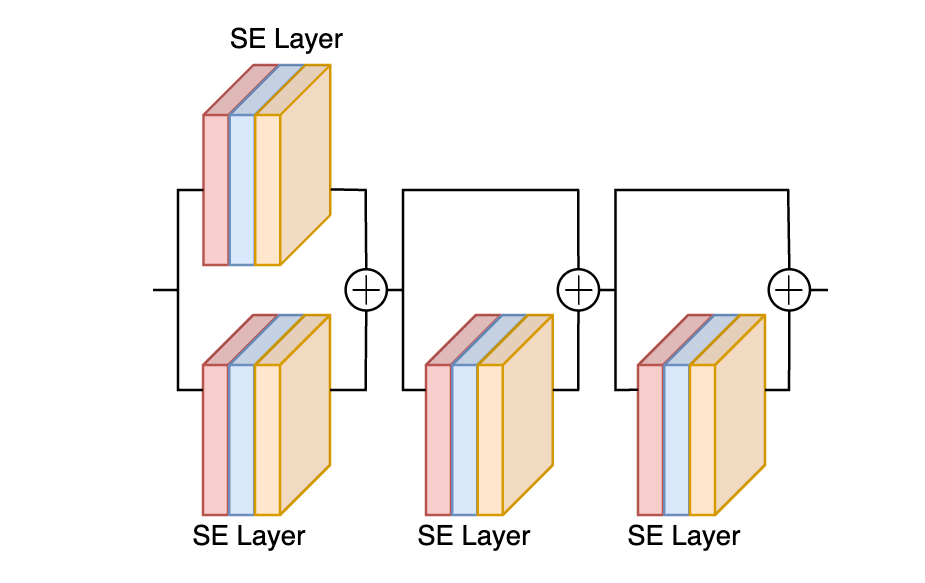}}&
{\includegraphics[width=0.3\textwidth]{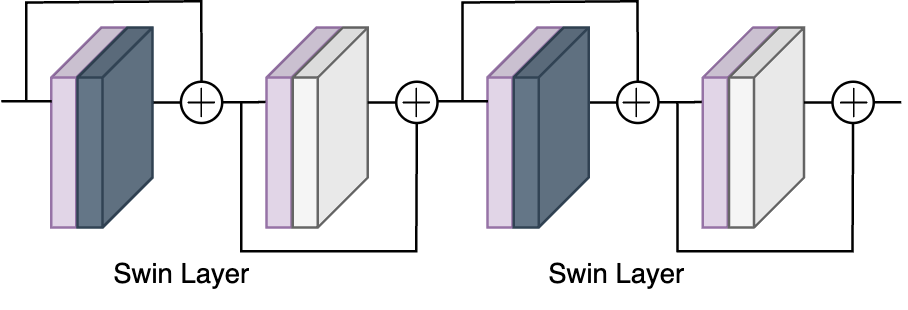}}&
{\includegraphics[width=0.3\textwidth]{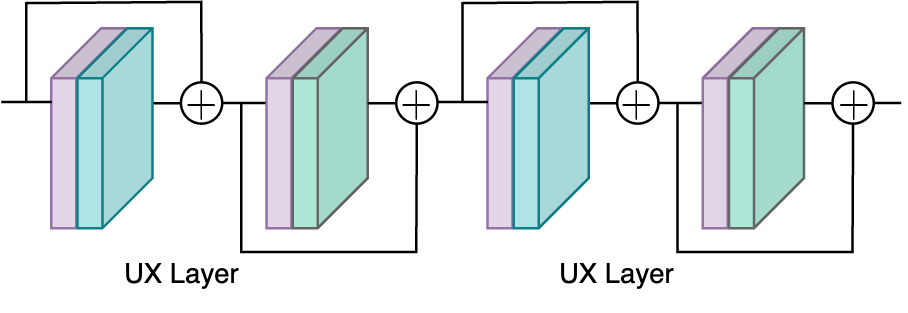}}
\\
(a)&(b)&(c)\\
\multicolumn{3}{c}{{\includegraphics[width=0.5\textwidth]{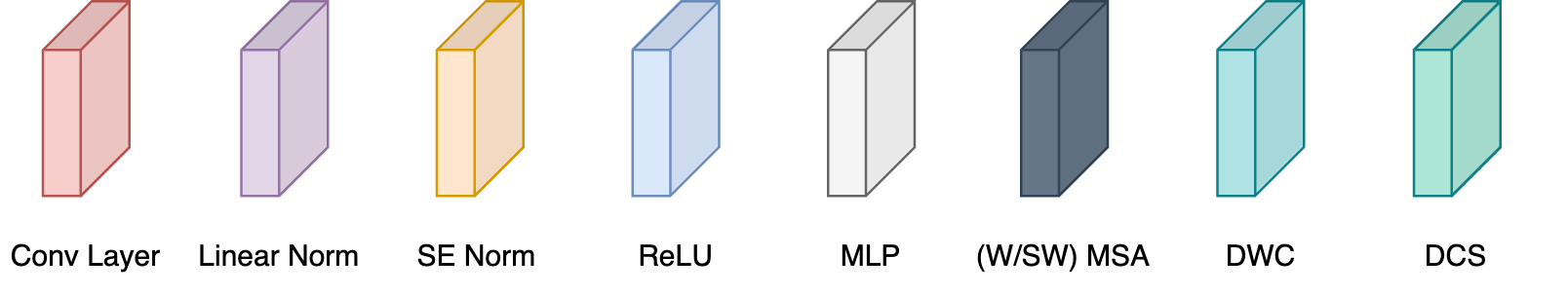}}}
\end{tabular}
\caption{The figure shows the inner structure of each block type used for our model construction. (a) is the squeeze-and-excitation block; (b) is the Swin block; and (c) is the UX block.}
\label{fig:blocks}
\end{figure*}

The internal mammary artery (IMA), aorta, and their surrounding perivascular adipose tissue (PVAT) have been recognized as clinically valuable in several studies as they have been shown to reflect inflammatory processes influencing cardiovascular health~\cite{kotanidis2022constructing,otsuka2013mammary,akoumianakis2019adipose}. The vascular wall secretes inflammatory molecules that diffuse into PVAT, triggering adipocyte changes at the perivascular level~\cite{oikonomou2019novel,kotanidis2022constructing,akoumianakis2019adipose}. In a recent study, Kotanidis et al.~\cite{kotanidis2022constructing} manually segmented these regions to assess the vascular inflammatory signature of COVID-19 (C19RS inflammatory signature) using CTA scans from 435 patients in the long-running Oxford Risk Factors and Non-Invasive Imaging (ORFAN) study. This novel non-invasive imaging biomarker, derived from the IMA, aorta, and PVAT, has demonstrated strong predictive power for acute vascular inflammation and in-hospital mortality. Additionally, it enables the extraction of reliable radiomic features from perivascular regions.

However, manual segmentation is highly time-consuming and labor-intensive, particularly as larger patient cohorts are required for improved generalizability. For instance, extending segmentation to new cohorts within the ORFAN study, which includes over 250,000 patient datasets, would make an automatic segmentation approach indispensable. Localizing the PVAT region is particularly challenging due to its small, suppressed appearance in axial views and its elongated, vertical structure in the chest. Therefore, this study focuses on developing an automated method for segmenting the IMA, aorta, and PVAT from CTA scans.

\begin{figure*}[t!]
\centering
{\includegraphics[width=\textwidth]
{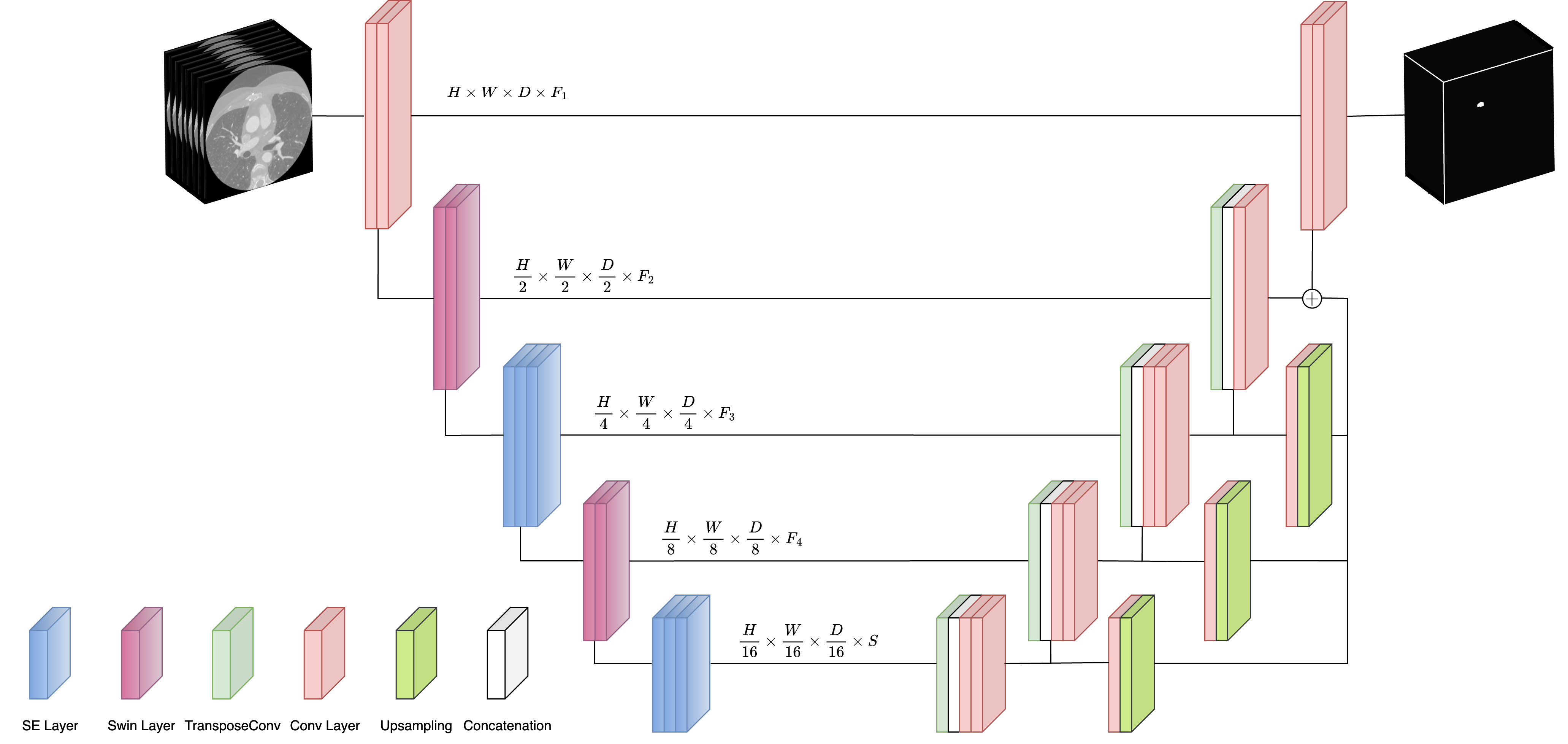}}
\caption{The figure shows the \emph{LegoNet} (specifically, \emph{LegoNet-2}) architecture. $F_{1-4}$ indicate the feature size, which is set to $\{24,48,96,192\}$, and $S$ is the hidden size, set to 768. This typical U-shaped architecture utilizes the block alternation concept, switching between Swin and SE blocks in the encoder in this example. The decoder is kept the same for all the variations of the model.}
\label{fig:model}
\end{figure*}

The key contributions of this work are as follows:
\begin{itemize}
    \item We introduce a novel deep learning paradigm that alternates different block types within a single architecture, demonstrating how the aggregation of diverse structural components enhances representation learning. The proposed \emph{LegoNet} achieves superior performance compared to state-of-the-art CNN and ViT-based models while maintaining lower complexity than ViT models.
    \item We address a previously unexplored problem in medical image analysis—the IMA and aorta PVAT space segmentation—which holds significant potential for cardiovascular disease prognosis and targeted therapeutic interventions.
    \item We conduct an extensive evaluation using external datasets, including intra-observer variability, inter-observer variability, model-versus-clinician performance analysis, and post-segmentation refinement studies with expert clinicians.
\end{itemize}

\section{Methodology}
We propose a simple yet effective alternating block method for constructing a DL architecture. Inspired by the modular nature of Lego blocks, this approach enables the integration of structurally diverse components to form a unified model, leveraging their complementary strengths to enhance feature representation and segmentation performance. Specifically, we explore three different types of blocks—CNN-based and SwinViT-based—and construct architectures that alternate between two of these blocks in various configurations.

\subsection{Building Blocks}
\subsubsection{SE block}
The squeeze-and-excitation (SE) block consists of stacks of a $3\times3\times3$ convolutional block with residuals, a ReLU activation function, and a SE normalization (norm) module~\cite{iantsen2021squeeze} within the layers, as shown in Figure~\ref{fig:blocks}(a). SE norm operates similarly to instance norm (IN)~\cite{ulyanov2016instance} but differs in the parameters $\gamma_i$ and $\beta_i$ in Equation~\ref{equation:senorm}. While IN treats these parameters as fixed during inference, SE norm dynamically models them as functions of the input, allowing for adaptive normalization based on feature responses~\cite{iantsen2021squeeze}

\begin{equation}
\label{equation:senorm}
    y_i = \gamma_ix^{\prime}_i+\beta_i,
\end{equation}
where $x^{\prime}_i$ is the normalized value of a batch of input data X, and $\gamma_i$ and $\beta_i$ are the scale and shift normalization values.

\subsubsection{Swin block}
Swin transformer~\cite{liu2021swin} with shifted windows has boosted the performance of ViT-based models due to its ability to capture global and local information. We employ the Swin block to see its compatibility with other CNN-based blocks and how well it performs in conjunction. The block consists of a linear normalization, regular and window partitioning multi-head attention (W-MSA and SW-MSA, respectively), and MLP, with skip connections as shown in Figure~\ref{fig:blocks}(b) and Equation~\ref{equation:swin}.

The outputs of the Swin block are computed in the sequential layers of $l$ and $l+1$ as:
\begin{equation*}
    \hat{z}^{l} = \text{W-MSA}(\text{LN}(z^{l-1})) + z^{l-1},
\end{equation*}
\begin{equation*}
    z^l = \text{MLP(LN}(\hat{z}^{l})) + \hat{z}^{l},
\end{equation*}
\begin{equation*}
    \hat{z}^{l+1} =\text{SW-MSA(LN}(z^l)) + z^l,
\end{equation*}
\begin{equation}
\label{equation:swin}
    z^{l+1} = \text{MLP(LN}(\hat{z}^{l+1})) + \hat{z}^{l+1},
\end{equation}

\noindent where $\hat{z}^l$ and $z^l$ are the outputs of the modules, W-MSA and SW-MSA denote regular and window partitioning multi-head self-attention modules, respectively, MLP is multi-layer perceptrons, and LN is a layer normalization.

\subsubsection{UX block} 
The UX block, introduced in~\cite{lee20223d}, is a convolution-based network block designed around large kernel sizes and depth-wise convolutions (DWC). Structurally, it mirrors the Swin block but replaces self-attention with depth-wise convolution using $7\times7\times7$ kernels, along with depth-wise convolutional scaling (DCS) and linear normalization as illustrated in Figure~\ref{fig:blocks}(c) and formulated in Equation~\ref{equation:ux}.

The outputs of the UX block are computed in the sequential layers of $l$ and $l+1$ as:

\begin{equation*}
    \hat{z}^{l} = \text{DWC}(\text{LN}(z^{l-1})) + z^{l-1}, 
\end{equation*}
\begin{equation*}
    z^l = \text{DCS(LN}(\hat{z}^{l})) + \hat{z}^{l},
\end{equation*}
\begin{equation*}
    \hat{z}^{l+1} =\text{DWC(LN}(z^l)) + z^l,
\end{equation*}
\begin{equation}
\label{equation:ux}
    z^{l+1} = \text{DCS(LN}(\hat{z}^{l+1})) + \hat{z}^{l+1},
\end{equation}

\noindent where $\hat{z}^l$ and $z^l$ are the outputs of the modules, DWC and DCS denote depthwise convolution (with kernel size starting from 7 × 7 × 7) and depthwise convolution scaling modules, respectively, and LN is a layer normalization.

\subsection{LegoNet Architecture}
The proposed network uses combinations of the blocks mentioned above. The input in the size of $X\in \mathbb{R}^{H\times W\times D\times C}$ (where $H$, $W$, $D$ and $C$ correspond to dimensions and the number of channels, respectively) passes through a stem block, as shown in Figure~\ref{fig:model}. This stem consists of two $3D$ convolutional blocks with $7\times7\times7$ and $3\times3\times3$ kernel sizes, respectively, rearranging the input to the size of $H\times W\times D\times 24$.

The alternating block mechanism is introduced at this stage, where two sets of blocks are applied in rotation. We propose three variations of this architecture, detailed in Section~\ref{section:variations}. Depicted in Figure~\ref{fig:model} is the second version of \emph{LegoNet} with Swin and SE blocks. The first block (i.e., Swin) downsamples the data to $\frac{H}{2}\times \frac{W}{2}\times \frac{D}{2}\times 48$. The next block (i.e., SE) reshapes the output to $\frac{H}{4}\times \frac{W}{4}\times \frac{D}{4}\times 96$. The same two blocks will repeat the procedure to generate the representations with the sizes $\frac{H}{8}\times \frac{W}{8}\times \frac{D}{8}\times 192$ and $\frac{H}{16}\times \frac{W}{16}\times \frac{D}{16}\times S$, respectively, where $S$ is the hidden size of the final block and is set to 768.

\subsection{Alternating Composition of LegoNet}
\label{section:variations}

Although we believe that \emph{LegoNet} as a concept is agnostic to the block type, we demonstrate the idea in three distinct versions, each differing in the block types used for model construction, as listed in Table~\ref{tab:configurations}. Figure~\ref{fig:model} illustrates the second version, alternating between Swin and SE blocks. The other versions follow the same structural framework, with SE and UX blocks in the first version and Swin and UX blocks in the third.

\begin{table}[htbp]
\centering
\caption{The table shows the different configurations for the network. These configurations can easily be changed in the code.}
\begin{tabular}{p{2cm}p{4cm}p{2cm}p{3cm}}    
\toprule    
    Network & Used blocks & Hidden size & Feature size  \\
    \midrule
    
    \emph{LegoNet-1} & SE$\rightarrow$UX$\rightarrow$SE$\rightarrow$UX & 768 & (24, 48, 96, 192) \\
    \emph{LegoNet-2} & Swin$\rightarrow$SE$\rightarrow$Swin$\rightarrow$SE & 768 & (24, 48, 96, 192) \\
    \emph{LegoNet-3} & Swin$\rightarrow$UX$\rightarrow$Swin$\rightarrow$UX & 768 & (24, 48, 96, 192) \\

    \bottomrule
\end{tabular}
\label{tab:configurations}
\end{table}

\subsection{Decoder}
The decoder of \emph{LegoNet} is designed to effectively integrate features from both the encoder output and the skip connections between the encoder and decoder. At each stage, encoder features are upsampled and concatenated with corresponding skip connection features, followed by two 3D convolutional blocks. The process repeats at each stage, with each block comprising upsampling, concatenation, and two convolutional blocks. At the same time, the outputs at each stage are carried over on which additional convolutional and upsampling are applied (See Figure~\ref{fig:model} right-most part). We perform this additional skip connection in the decoder to leverage a better flow of features during reconstruction. The final segmentation head constitutes two 3D convolutional blocks to generate the segmentation masks.

\subsection{Model Refinement via Iterative Learning}

\begin{figure}[hbt!]
    \centering
    \includegraphics[width=\textwidth]{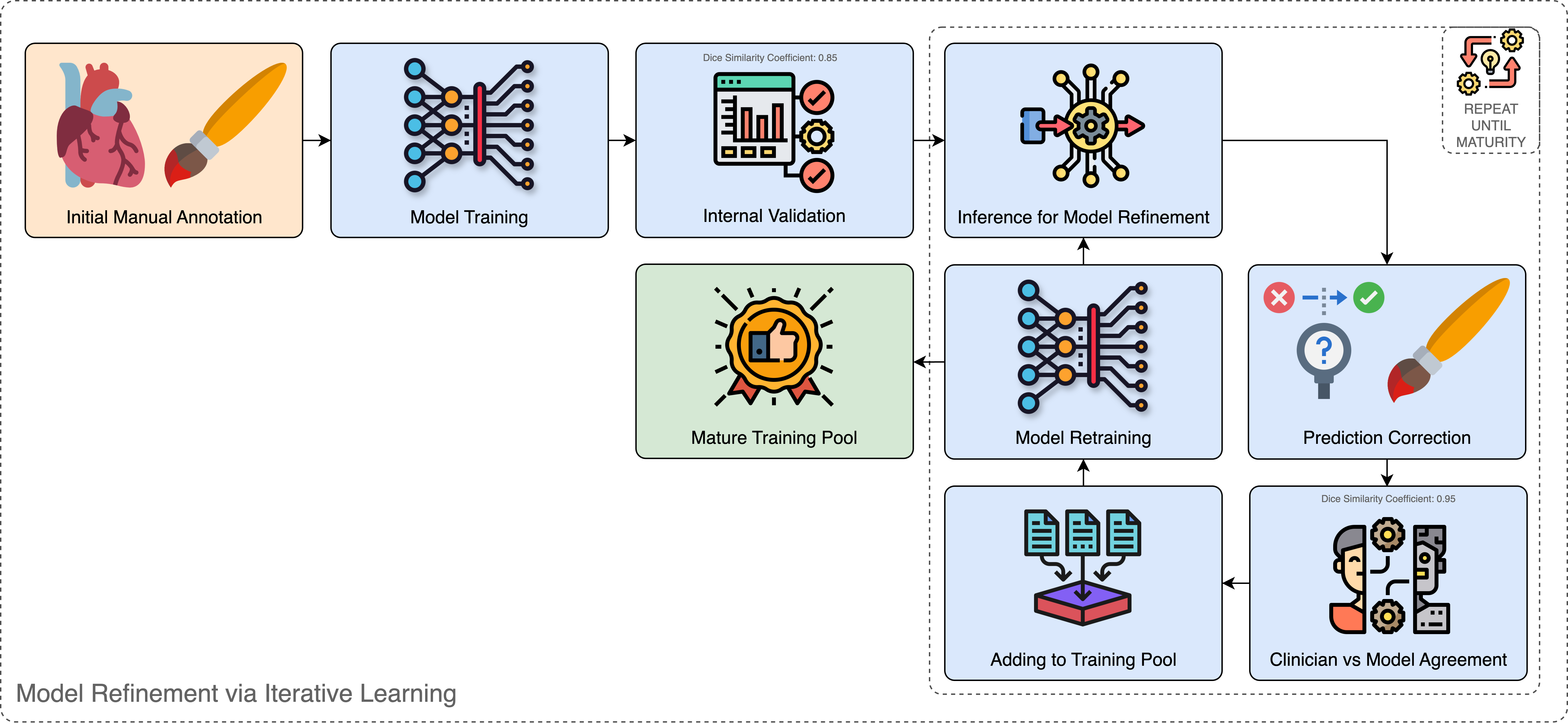}
    \caption{Model Refinement via Iterative Learning. This approach improves the segmentation model to maturity before deploying it in large cohort data. The model is initially trained with a small, feasible cohort and is internally validated. Several cohorts of data are then used to improve the model by adding more value to the learning process iteratively.}
    \label{fig:active_learning}
\end{figure}

Given the high cost of manual vascular segmentation, we employ a cost-effective approach called model refinement via iterative learning, which enables model development with a limited dataset while progressively improving performance until maturity. Before applying the deep learning model to large cohorts like ORFAN, we ensure it reaches maturity, which we quantify through two key factors: (i) model refinement through iterative learning, ensuring high segmentation performance and (ii) inter- and intra-observer variability analysis, validating consistency against expert clinicians. 

Guided by clinical feedback, we implement model refinement via iterative learning, as illustrated in Figure~\ref{fig:active_learning}. The process begins by training the model on a small, ground-truth-labeled dataset, followed by internal validation using k-fold cross-validation. Next, we run inference on a new batch of data, and clinicians correct the model's predictions, significantly reducing the manual effort compared to annotating from scratch. The corrected masks are then compared against the model's predictions and incorporated into the training pool to refine the model further. This iterative process continues until a highly acceptable performance is reached (e.g., DSC > 0.90).

Additionally, we conduct a second set of experiments to assess the model's reliability against inter- and intra-observer variability. In this study, a separate small dataset is blindly annotated by one clinician at two different time points and by another clinician once. The model's performance is then evaluated against the agreement between clinicians, serving as an external benchmark for segmentation accuracy and robustness.

\section{Dataset and Preprocessing}
The proposed concept was trained and validated using a multi-cohort, multi-scanner subset of ORFAN and the publicly available ASOCA~\cite{gharleghi2022automated,gharleghi2023annotated} dataset. Our dataset comprises 155 patients from three different centres for initial model training and validation, followed by 49 patients designated for inter- and intra-observer variability analysis. We used three additional cohorts comprising 54, 41, and 39 patients, all sourced from different UK sites for ``model refinement via iterative learning'' process. Furthermore, we incorporate an additional subset from the U.S., comprising 712 scans. All of these cohorts are sourced from ORFAN, and detailed information on data acquisition and study protocols can be found in~\cite{chan2024inflammatory,kotanidis2022constructing}. Finally, we used the publicly available ASOCA dataset~\cite{gharleghi2022automated,gharleghi2023annotated} for external validation, which includes 30 healthy subjects and 30 patients diagnosed with coronary artery disease, to test the model's performance in a different study protocol.

Manual segmentation is performed around the IMA, which extends from the level of the aortic arch to 120mm caudally. One diameter of the IMA defines the perivascular space. In contrast, the aorta is segmented from the bifurcation point, extending 67.5mm caudally. Its perivascular adipose tissue (PVAT) is similarly measured as one diameter of the aorta.

Since the datasets originate from multiple centres, variations exist in scanning parameters, scanner types, and image characteristics, leading to differences in scan dimensions, spacing, orientation, and direction. We apply a standardised preprocessing pipeline to ensure consistency, aligning all scans to a uniform direction and orientation with isotropic spacing of $1\times1\times1 mm^3$. Additionally, we clip CT intensity values to the range [-1024, 1024] and normalize them to [-1, 1] for improved numerical stability and model robustness.


\section{Experimental Setup}
We evaluate our proposed method against a range of state-of-the-art deep learning networks, including U-Net~\cite{kerfoot2019left}, SegResNet~\cite{myronenko20193d}, UNETR~\cite{hatamizadeh2022unetr}, Swin UNETR~\cite{hatamizadeh2022swin}, UX-Net~\cite{lee20223d}, and UNesT~\cite{yu2022unest}. These models are first rigorously tested on the IMA+PVAT segmentation task, followed by an extended evaluation on the aorta+PVAT task for further comparison. All models are trained for 100 epochs, starting from random initialization. 

For training, we use the AdamW optimizer with a learning rate of $1e-3$, weight decay of $1e-5$, and cosine annealing scheduler with minimum $\eta$ of $1e-5$ and $T_0$ at 25. The batch size is set to 1, and the loss function is computed as the sum of Dice and Focal losses (Equation~\ref{dice_loss} and Equation~\ref{focal_loss}) for segmentation. All experiments are conducted on a single NVIDIA Tesla V100 GPU.
 
\begin{equation}
   \label{dice_loss}
   \mathcal{L}_{Dice} = \frac{2\sum_{i}^{N} \hat{y_i} y_i}{\sum_{i}^{N} \hat{y_i}^2 + \sum_{i}^{N} y_i^2},
\end{equation}

\begin{equation}
   \label{focal_loss}
   \mathcal{L}_{Focal} = -\sum_{i}^{N}\epsilon y_i (1 - \hat{y_i})^{\psi}log(\hat{y_i}) - (1 - y_i)\hat{y_i}^{\psi}log(1-\hat{y_i}),
\end{equation}

\begin{equation}
    \label{final_loss}
   \mathcal{L}_{Segmentation} = \mathcal{L}_{Dice} + \mathcal{L}_{Focal}
\end{equation}

\noindent where $\hat{y}$ is the prediction of the model, $y$ is the ground truth, $\epsilon$ is the weightage for the trade-off between precision and recall in the focal loss (empirically set to 1), $\psi$ is focusing parameter (set to 2), and $N$ is the sample size.

The primary performance metric for evaluation is the Dice Similarity Coefficient (DSC). Additionally, we report precision, recall, and the 95\% Hausdorff Distance to provide a more comprehensive comparison. The results are presented as the mean and standard deviation from 5-fold cross-validation on the training and validation data. We compare the number of learnable parameters and floating-point operations (FLOPs) for each model to assess model complexity. DSC and volume-based comparisons are further analysed in the clinical evaluation section to assess segmentation performance in a real-world clinical setting.

\section{Results}

\begin{table*}[t!]
\scriptsize
\centering
\caption{The table reports the mean and standard deviation of DSC, precision, recall, and HD95 for 5-fold cross-validation and the number of parameters and FLOPs of different models. All the experiments in this table are trained with random initialization.}
\begin{tabular}{p{2cm}p{1.5cm}p{1.5cm}p{1.5cm}p{1.5cm}p{1.5cm}p{1.5cm}}    

    \toprule
    Models & DSC$\uparrow$ & Precision$\uparrow$ & Recall$\uparrow$ & HD95$\downarrow$ & Params (M) & FLOPs (G) \\
    \midrule
    UNet~\cite{cciccek20163d,kerfoot2019left} & 0.686$\pm$0.03 & 0.72$\pm$0.04 & 0.69$\pm$0.03 & 2.70 & 3.99 & 27.64 \\
    SegResNet~\cite{myronenko20193d} & 0.732$\pm$0.01 & 0.75$\pm$0.02 & 0.74$\pm$0.03 & 2.50 & 4.7 & 61.71 \\
    UX-Net~\cite{lee20223d} & 0.695$\pm$0.03 & 0.73$\pm$0.06 & 0.70$\pm$0.01 & 3.17 & 27.98 & 164.17 \\
    \midrule
    UNETR~\cite{hatamizadeh2022unetr} & 0.690$\pm$0.02 & 0.72$\pm$0.03 & 0.69$\pm$0.03 & 3.00 & 92.78 & 82.48 \\
    SwinUNETR~\cite{hatamizadeh2022swin} & 0.713$\pm$0.02 & 0.74$\pm$0.02 & 0.71$\pm$0.04 & 2.46 & 62.83 & 384.20 \\
    UNesT~\cite{yu2022unest} & 0.555$\pm$0.04 & 0.59$\pm$0.06 & 0.55$\pm$0.05 & 4.35 & 87.20 & 257.91 \\
    \midrule
    \emph{LegoNet-1} & 0.747$\pm$0.02 & 0.75$\pm$0.02 & \textbf{0.77}$\pm$0.03 & 2.34 & 50.58 & 175.77 \\
    \emph{LegoNet-2} & \textbf{0.749}$\pm$0.02 & \textbf{0.77}$\pm$0.01 & 0.76$\pm$0.04 & \textbf{2.11} & 50.71 & 188.02 \\
    \emph{LegoNet-3} & 0.741$\pm$0.02 & 0.76$\pm$0.02 & 0.75$\pm$0.03 & 2.34 & 11.14 & 173.41 \\
    \bottomrule
    
\end{tabular}
\label{tab:main_results}
\end{table*}

\subsection{Initial Model Training} 
For efficiency purposes, we investigated the performance of different architectures on the initial dataset of 155 scans in the IMA+PVAT task. Table~\ref{tab:main_results} presents the segmentation performance and model complexities.

Among the baseline models, U-Net (CNN-based) and UNETR (ViT-based) exhibit similar performance, with mean DSC scores of 0.686 and 0.690, respectively. UX-Net achieves a slightly higher DSC of 0.695, while UNesT significantly underperforms with a DSC of 0.555. SwinUNETR shows a notable improvement, yielding a DSC of 0.713, whereas SegResNet demonstrates the highest performance among existing models.

All three variations of LegoNet surpass the baseline models across DSC, precision, recall, and HD95 metrics. LegoNet-2 (Swin and SE alternation) achieves the highest DSC of 0.749, followed closely by the other two versions with DSC scores of 0.747 and 0.741, respectively. A similar trend is observed across precision, recall, and HD95, with LegoNet consistently outperforming existing architectures.

\subsection{Statistical Analysis}
To further assess the model's performance, we performed a statistical significance analysis comparing \emph{LegoNet} with SegResNet and SwinUNETR, the two strongest baseline models. This analysis is based on the results of the initial data set presented in Table~\ref{tab:main_results}.

We apply the Wilcoxon signed rank test~\cite{wilcoxon1992individual} to determine whether \emph{LegoNet} exhibits statistically significant improvements over competing models. The null hypothesis $H_0$ assumes that the segmentation performance of \emph{LegoNet} is statistically indistinguishable from the other models, while the alternative hypothesis $H_1$ posits that \emph{LegoNet} outperforms SwinUNETR and SegResNet.

The results of the Wilcoxon signed rank test reveal a $p$ value of 1.59e-4 for the \emph{LegoNet} vs. SegResNet comparison and a $p$ value of 2.13e-10 for \emph{LegoNet} vs. SwinUNETR, both indicating highly significant differences. These findings confirm that \emph{LegoNet} is not only the best-performing model in terms of DSC but also statistically superior to the strongest baselines.

\subsection{Clinical Evaluation Setting}
Once LegoNet was cross-validated, we evaluated its performance in a clinical setting through two key analyses: (i) inter-/intra-observer variability analysis and (ii) post-model agreement analysis.

\subsubsection{Inter- and Intra-observer Variability Analysis.}

We conducted a comparative segmentation study on a new cohort of 49 scans to evaluate the model's agreement with human experts. Two expert clinicians performed manual segmentation and we compared their annotations with the automatic segmentations generated by LegoNet. For intraobserver variability, an expert radiologist with six years of experience manually segmented the same cohort twice, with a 12-month interval between annotations. DSC between these two instances was 0.804, reflecting intra-rater consistency. For interobserver variability, a less senior radiologist with three years of experience independently segmented the same cohort. The inter-clinician variability, calculated as the DSC between the two manual segmentations from different clinicians, reached 0.761.

We computed the mean DSC between the model's segmentations and the three manual annotations (two from the first clinician and one from the second) to assess model vs. human agreement. The model vs. human agreement resulted in a DSC of 0.733, demonstrating strong alignment with expert annotations. 





\begin{table}[t!] 
\scriptsize
\centering
\caption{DSC, recall, precision metrics for a random split in the aorta segmentation. The same models were validated with the same settings as IMA+PVAT. The proposed model variations performed consistently with a different but relatively easier task of aorta segmentation.}
    \begin{tabularx}{0.6\columnwidth}{@{\extracolsep{\fill}} lccc }
        \toprule
        Models & DSC$\uparrow$ & Precision$\uparrow$ & Recall$\uparrow$ \\
        \midrule
        UNet & 0.895 & 0.907 & 0.891 \\
        SegResNet & 0.885 &  0.875  & 0.900 \\
        UX-Net & 0.919 & 0.918 & 0.925 \\
        \midrule
        UNETR & 0.817 & 0.827 & 0.831 \\
        SwinUNETR & 0.906 & 0.887 & 0.931 \\
        UNesT & 0.838 & 0.853 & 0.847 \\
        \midrule
        \emph{LegoNet-1} & \textbf{0.939} & \textbf{0.919} & \textbf{0.961} \\
        \emph{LegoNet-2} & 0.898 & 0.850 & 0.957 \\
        \emph{LegoNet-3} & 0.903 & 0.912 & 0.891\\
        \bottomrule

    \end{tabularx}
    \label{tab:aorta_results}
\end{table}

\subsubsection{Post-Model Agreement Analysis via Iterative Refinement.}
We conducted a post-model agreement analysis (see Figure~\ref{fig:active_learning}) using an iterative learning strategy to improve model performance and assess its adaptability to new cohorts. We generated segmentation masks for three completely unseen cohorts ($n=54$, $n=41$, and $n=39$) (distinct from training, validation, and inter/intra-observer datasets), and a clinician corrected the model predictions. These refined segmentations were added to the training set, increasing dataset diversity and improving model performance. This process was repeated three times, progressively expanding the dataset.

\subsubsection{Volume-Based Analysis and Model Refinement Impact}
\begin{figure*}[t!]
\centering
\begin{tabular}{ccc}
{\includegraphics[width=0.32\textwidth]{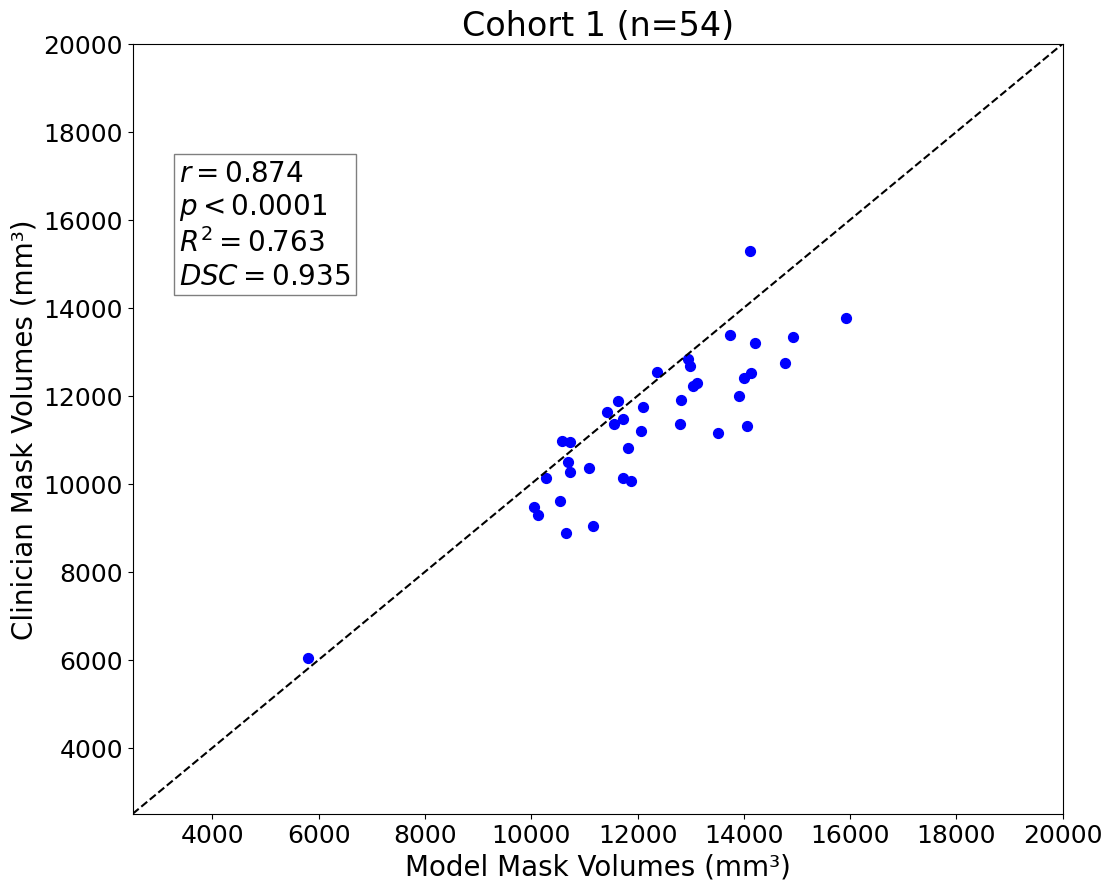}}&
{\includegraphics[width=0.32\textwidth]{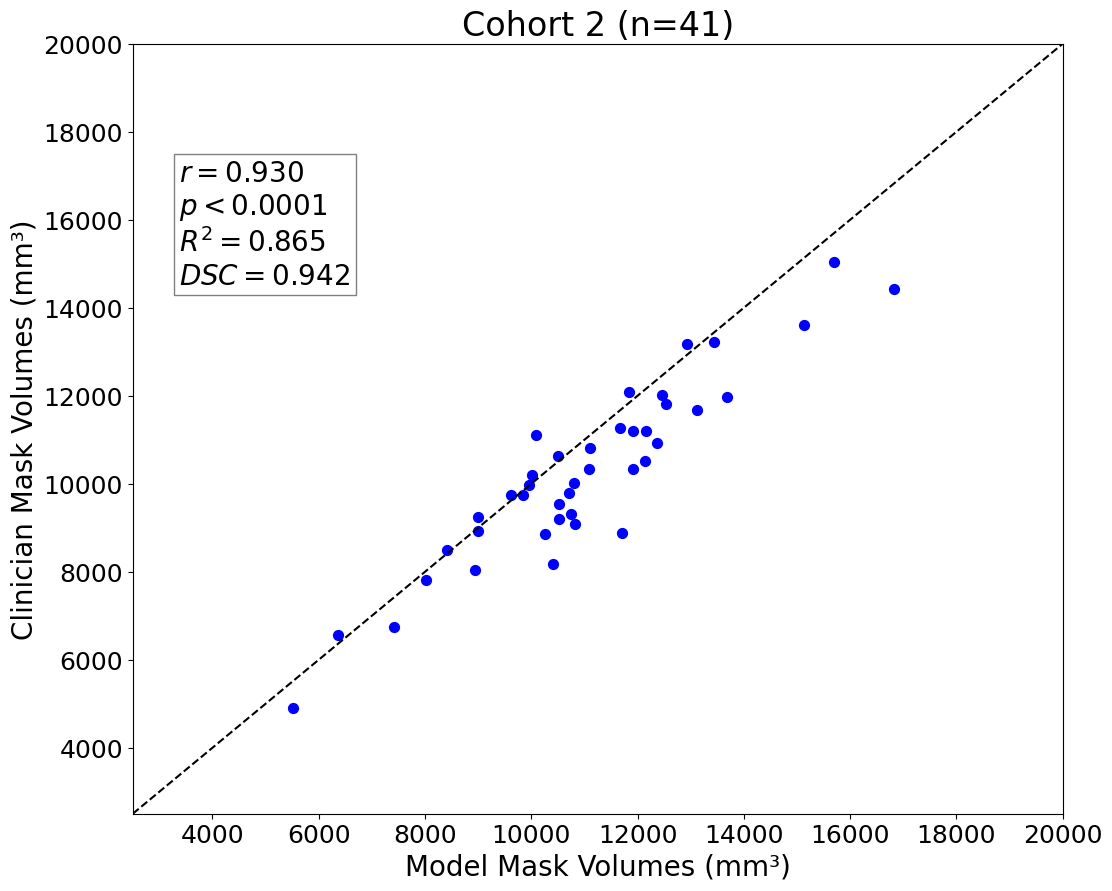}}&
{\includegraphics[width=0.32\textwidth]{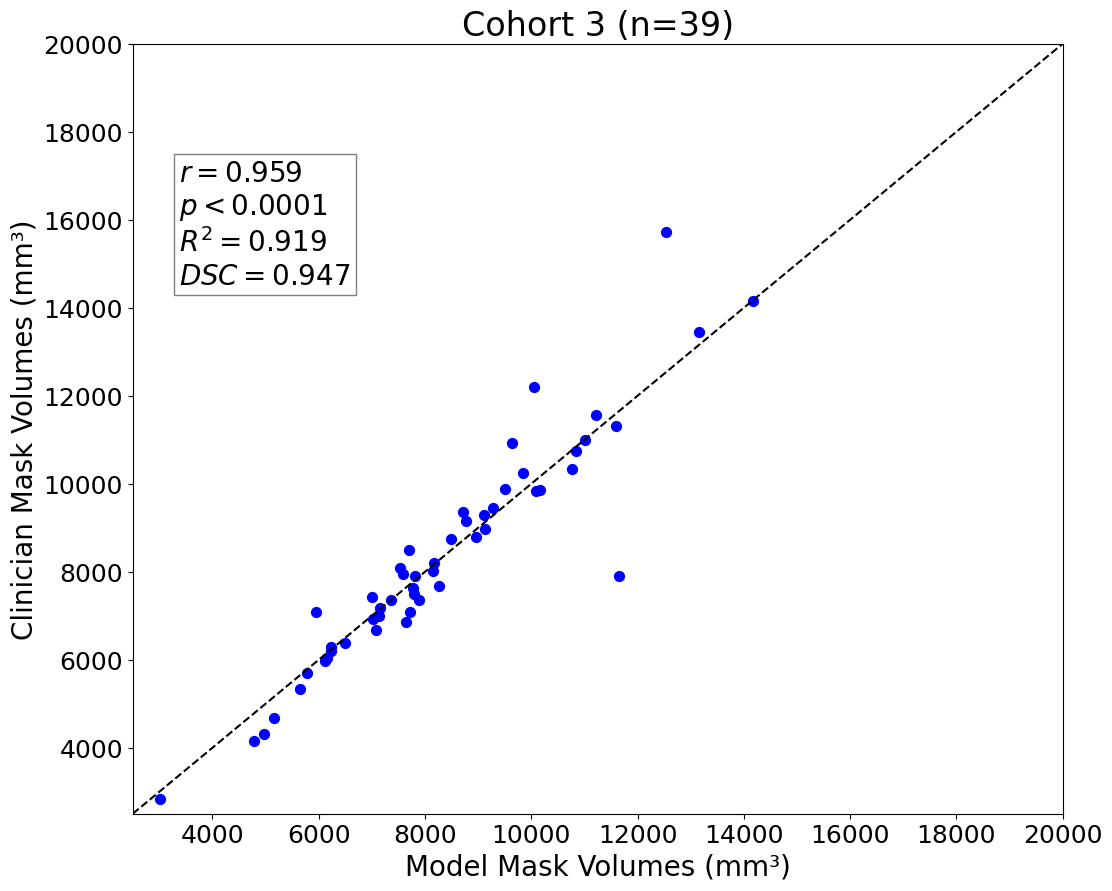}}
\\
{\includegraphics[width=0.32\textwidth]{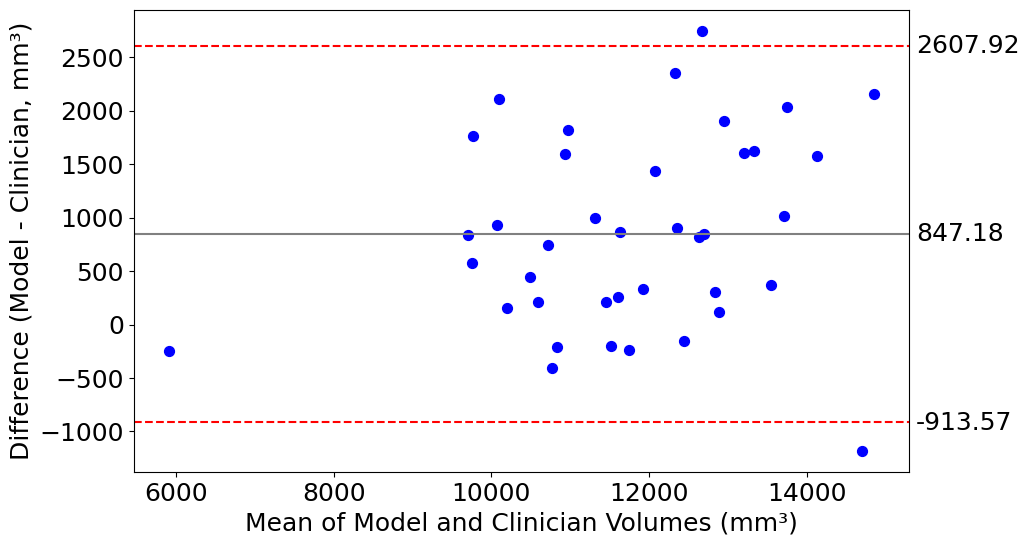}}&
{\includegraphics[width=0.32\textwidth]{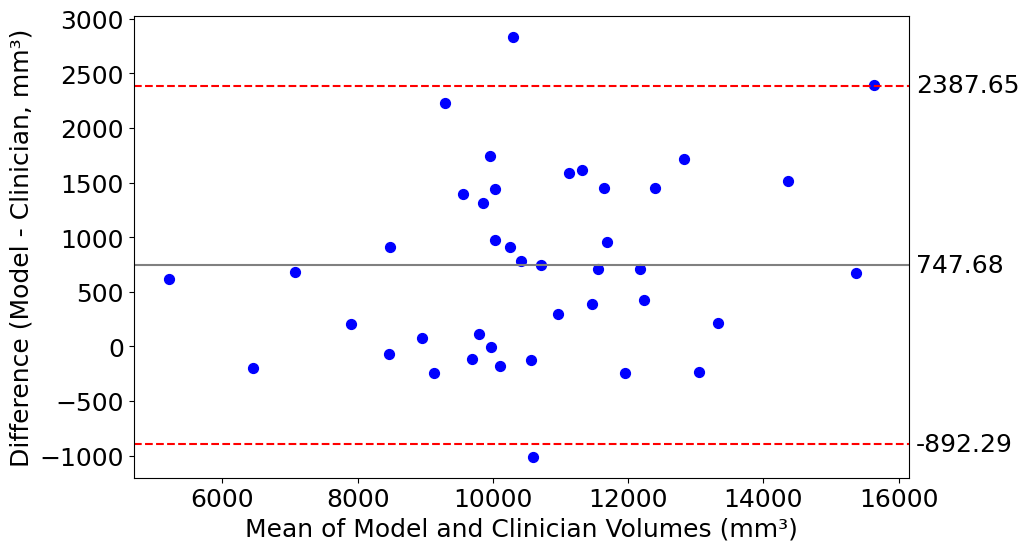}}&
{\includegraphics[width=0.32\textwidth]{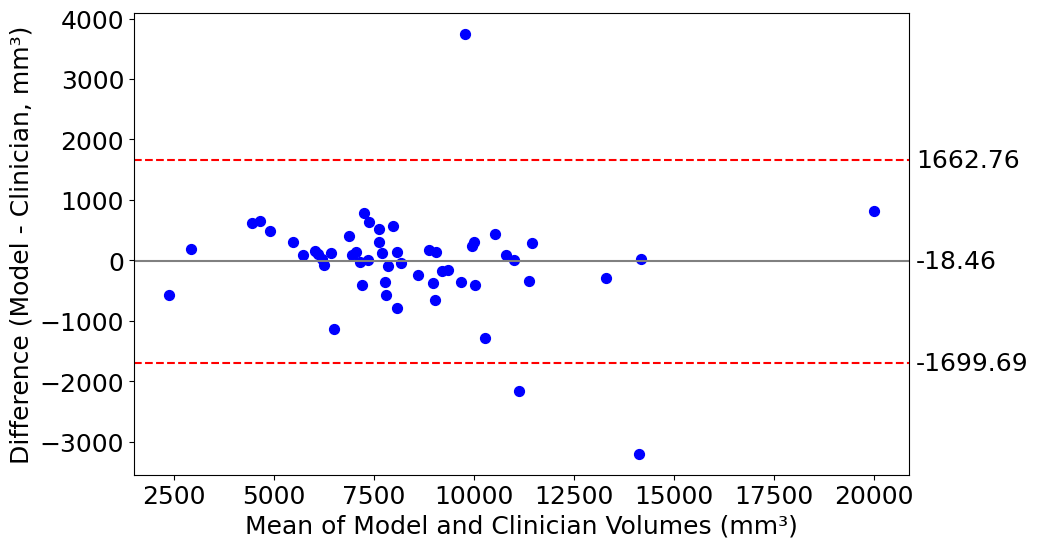}}
\\
(a) Cohort 1 ($n=54$) &(b) Cohort 2 ($n=41$) &(c) Cohort 3 ($n=39$)
\end{tabular}
\caption{The figure shows the correlation and Bland-Altman plots for three external cohorts, comparing the model's prediction and clinician's segmentation masks.}
\label{fig:volumes}
\end{figure*}

In Figure~\ref{fig:volumes}, we present a volume-based comparison of segmentation performance in different refinement stages. We computed the segmentation volume for each patient using the clinician's manual annotations and LegoNet's automatic predictions. In the first cohort (Figure~\ref{fig:volumes}(a)), the model over-segmented the IMA \& perivascular space for many patients, with a Mean Absolute Error (MAE) of 0.982, Spearman's $\rho$ of 0.874 ($p$<0.0001), and DSC of 0.935. With iterative refinement, the segmentation accuracy progressively improved. By the third cohort (Figure~\ref{fig:volumes}(c)), the model's predictions closely matched the clinician's annotations, with an MAE of 0.491, Spearman's $\rho$ of 0.959 ($p$<0.0001), and DSC of 0.947, demonstrating effective learning from corrections. These findings underscore the progressive enhancement of model performance through iterative refinement.

\subsection{Evaluation on a Large External U.S. Cohort}
To further assess the generalizability of LegoNet, we retrained the model on the entire dataset. We examined its performance on a completely new US cohort consisting of 712 scans (part of the ORFAN study). Each predicted segmentation mask was reviewed by an expert clinician for the sole purpose of quality assurance. 32 cases were rejected due to a limited field of view (FOV), where either the IMA or aorta (or both) were partially or entirely outside the scan range. The remaining 680 scans were deemed clinically acceptable and were subsequently used for downstream tasks.

\subsection{External Public Data}
The primary objective of evaluating LegoNet in the public ASOCA cohort is to demonstrate its generalisability in (i) different acquisition techniques, (ii) different imaging machines and centres, and (iii) different medical protocols. Unlike internal data sets, the ASOCA cohort was collected under different medical protocols, providing a challenging validation scenario for the model.

For this evaluation, we again used the model, which had been trained with all in-house CTA data, including the training/validation set and external in-house cohorts, totalling 338 patient cases. The expert clinician manually corrected the model's segmentation masks, and we computed the DSC agreement. Remarkably, the model maintains high consistency with the three previously tested model refinement cohorts, achieving a DSC of 0.961, a precision of 0.961, and a recall of 0.938. To support further research and validation, the segmentation masks for this cohort will be made publicly available upon request.

\subsection{Evaluation on aorta}
All the experiments discussed above were conducted on the IMA+PVAT segmentation task. To further assess the generalizability of the proposed approach, we extended the study to evaluate the same models on the aorta+PVAT segmentation task. LegoNet consistently outperformed other leading architectures across DSC, precision, and recall, maintaining superior performance across different segmentation tasks (see Table~\ref{tab:aorta_results}). In the aorta segmentation task, the UX-Net and SwinUNETR achieve better performance compared to other CNN and ViT models, with 0.919 and 0.906 DSC, respectively. Version one of LegoNet reaches the highest performance with 0.939 DSC, 0.919 precision, and 0.961 recall values. While the other two versions are slightly lower, they are on par with other leading architectures. These findings suggest that LegoNet is robust and generalizable, effectively adapting to similar vascular segmentation problems. 

\section{Discussion}
This study addresses a novel medical imaging challenge - the automatic segmentation of the IMA, aorta, and PVAT from CTA images. This segmentation is a critical precursor to predictive prognostic modelling, facilitating risk assessment and patient outcome prediction in subsequent studies. The clinical value of these segmented regions has already been demonstrated in predicting acute vascular inflammation and in-hospital mortality~\cite{kotanidis2022constructing}. The PVAT analysis is not limited to acute inflammation and can and will be extended to capture other molecular changes in the region, such as fibrosis, adipogenesis, lipolysis, etc. This investigation can eventually lead to a better understanding of the molecular mechanisms driving these medical disorders, unlocking avenues to new therapeutic targets.

To tackle this problem, we introduce a new deep learning paradigm based on block alternation, where structurally distinct yet complementary blocks are interleaved to enhance feature learning. We propose three variations of LegoNet, all of which outperform leading CNN- and ViT-based models on multi-centre datasets. Additionally, we examined the models' complexities to ensure the balance of performance and cost. Finally,  the proposed model is exhaustively tested in multiple settings and cohorts.

We observe a discrepancy between the cross-validation results ($\approx$0.750 DSC) and post-model agreement on external cohorts ($\approx$0.900 DSC). This difference is primarily attributed to variability in segmentation interpretation. In clinical practice, expert clinicians accept model-generated masks as valid representations of the IMA and perivascular space, provided they are sufficiently accurate for diagnostic purposes~\cite{kotanidis2022constructing}. Our inter- and intra-observer variability and model vs. human agreement analyses further confirm that these results align with expected variability in manual segmentation.

We attribute the superior performance of \emph{LegoNet} to (i) structurally different blocks that are assumed to learn more discriminative features and (ii) the complexity of the model. Compared to CNN models, the complexity in the number of parameters and GFLOPs is much higher. However, that is on par with ViT models, such as UNETR, SwinUNETR, and UNesT. The best-performing \emph{LegoNet-2}, for example, stands at 50.71M parameters and 188.02G FLOPs, which is smaller than the three ViT-driven models. In the future, the model's behaviour with more recent models, such as Mamba-based blocks, could be studied. The use of more than two repetitive blocks could be another avenue to investigate.

\section{Conclusion}
This work introduces a new deep learning paradigm that alternates structurally distinct blocks, leveraging their complementary strengths to construct a more effective architecture. Moving beyond the conventional approach of using identical blocks, we demonstrate that integrating dissimilar blocks enhances model learning. LegoNet consistently outperforms leading CNN and ViT-based models on two CTA datasets, with further validation on external, international, and public cohorts, where clinician-model agreement in DSC remains high. Additionally, intra- and inter-observer variability studies further confirm the reliability of our approach. We propose three variations of LegoNet, applying this concept to segment the IMA, aorta, and their perivascular space — a clinically valuable but previously unstudied region. Given its proven significance in vascular inflammation and cardiovascular disease prognosis, accurate segmentation of PVAT regions holds potential for advancing risk assessment and therapeutic planning.

%
%
\bibliographystyle{splncs04}
\bibliography{bib}
\end{document}